\documentclass[9pt,twocolumn,twoside]{osajnl}

\usepackage{braket}

\journal{ol} 

\setboolean{shortarticle}{true}

\title{Analysis of intensity correlation enhanced plasmonic structured illumination microscopy}

\author[1,2,3,*]{Anton Classen}
\author[1,2]{Xinghua Liu}
\author[1,4,5,6]{Aleksei M. Zheltikov}
\author[1,2,3]{Girish S. Agarwal}

\affil[1]{Institute for Quantum Science and Engineering, Texas A\&M University, College Station, TX 77843-4242, USA}
\affil[2]{Department of Physics and Astronomy, Texas A\&M University, College Station, TX 77843-4242, USA}
\affil[3]{Department of Biological and Agricultural Engineering, Texas A\&M University, College Station, TX 77843-4242, USA}
\affil[4]{Physics Department, International Laser Center, M. V. Lomonosov Moscow State University, Moscow 119992, Russia}
\affil[5]{Russian Quantum Center, Skolkovo, Moscow Region 143025, Russia}
\affil[6]{Kazan Quantum Center, A. N. Tupolev Kazan National Research Technical University, Kazan 420126, Russia}

\affil[*]{Corresponding author: aclassen@tamu.edu}

\begin{abstract}
We propose to enhance the performance of localized plasmon structured illumination microscopy (LP-SIM) via intensity correlations. LP-SIM uses sub-wavelength illumination patterns to encode high spatial frequency information. It can enhance the resolution up to three-fold before gaps in the OTF support arise. For blinking fluorophores or for quantum antibunching an intensity correlation analysis induces higher harmonics of the illumination pattern and enlarges the effective OTF. This enables ultrahigh resolutions without gaps in the OTF support, and thus a fully deterministic imaging scheme. We present simulations that include shot and external noise and demonstrate the resolution power under realistic photon budgets. The technique has potential in light microscopy where low-intensity illumination is paramount while aiming for high spatial but moderate temporal resolutions.  
\end{abstract}

\setboolean{displaycopyright}{true}

\begin{document}

\maketitle

The resolution of optical microscopy is limited by diffraction to around half the wavelength of light. Today, various superresolution microscopy techniques overcome this limit by drawing on effects from statistical, non-linear and quantum optics \cite{Gustafsson2000, Hell1994, Betzig2006, Dertinger2009, Schwartz2013}. Yet, each method comes with specific requirements and trade-offs, leading to either low imaging speeds, limited resolution improvements, high complexity of use, limited applicability, or the need for prohibitively high illumination intensities. Structured illumination microscopy (SIM) is considered to be one of the most powerful and versatile techniques due to its combination of resolution improvement with good acquisition speed and flexibility of use \cite{Strohl2016}. SIM relies on spatial frequency mixing to double the resolution within the linear regime. Another approach utilizes intensity auto-correlations, evaluated from an image series. For these intensity correlation microscopy (ICM) techniques, statistically blinking fluorophores \cite{Dertinger2009} or quantum emitters that exhibit anti-bunching \cite{Schwartz2013} can be used. The first approach is well-known as super-resolution optical fluctuation imaging (SOFI) \cite{Dertinger2009}. SOFI can be implemented without adjustments to microscope system but it leverages temporal resolution to achieve a higher spatial resolutions as it is necessary to acquire sufficient blinking statistics within a frame set.  

Recently, we found that SIM and ICM can be combined into \textit{structured illumination intensity correlation microscopy} (SI-ICM) \cite{Classen2017a,Classen2018}. For correlation order $m$ the enhancement scales favorably as up to $m + m = 2m$. Since ICM and SIM are both linear, SI-ICM can operate at linear low-intensity illuminations. We point out that the SI-ICM principle can equally be applied in confocal microscopy as was successfully demonstrated using  anti-bunching \cite{Tenne2019} or blinking \cite{Sroda2020}. The advantage of widefield SIM --- as compared to confocal --- is the stronger amplification of high spatial frequency information, resulting in better effective resolutions \cite{Strohl2016}. This is the result of selective spatial frequency mixing with only the highest spatial frequency $\pm \mathbf{k}_0 $ of the illumination pattern $I_{\text{str}}(\mathbf{r}) = 1 + 1 \cos(\mathbf{k}_0\mathbf{r}+\varphi)$. Linear SIM is diffraction-limited to illumination frequencies $\mathbf{k}_{0} \leq \mathbf{k}_\text{max}$, leading to the total passband $|\mathbf{k}_{0} + \mathbf{k}_\text{max}| \leq 2 k_\text{max}$, where $k_\text{max}$ is the highest spatial frequency transmitted by optical transfer function (OTF) of the imaging system $H(\mathbf{k}) \equiv FT  \{ h(\mathbf{r}) \}$, and $h(\mathbf{r})$ is the corresponding point spread function (PSF).

\begin{figure}[t]
  \centering
  \includegraphics[width=1.0 \columnwidth]{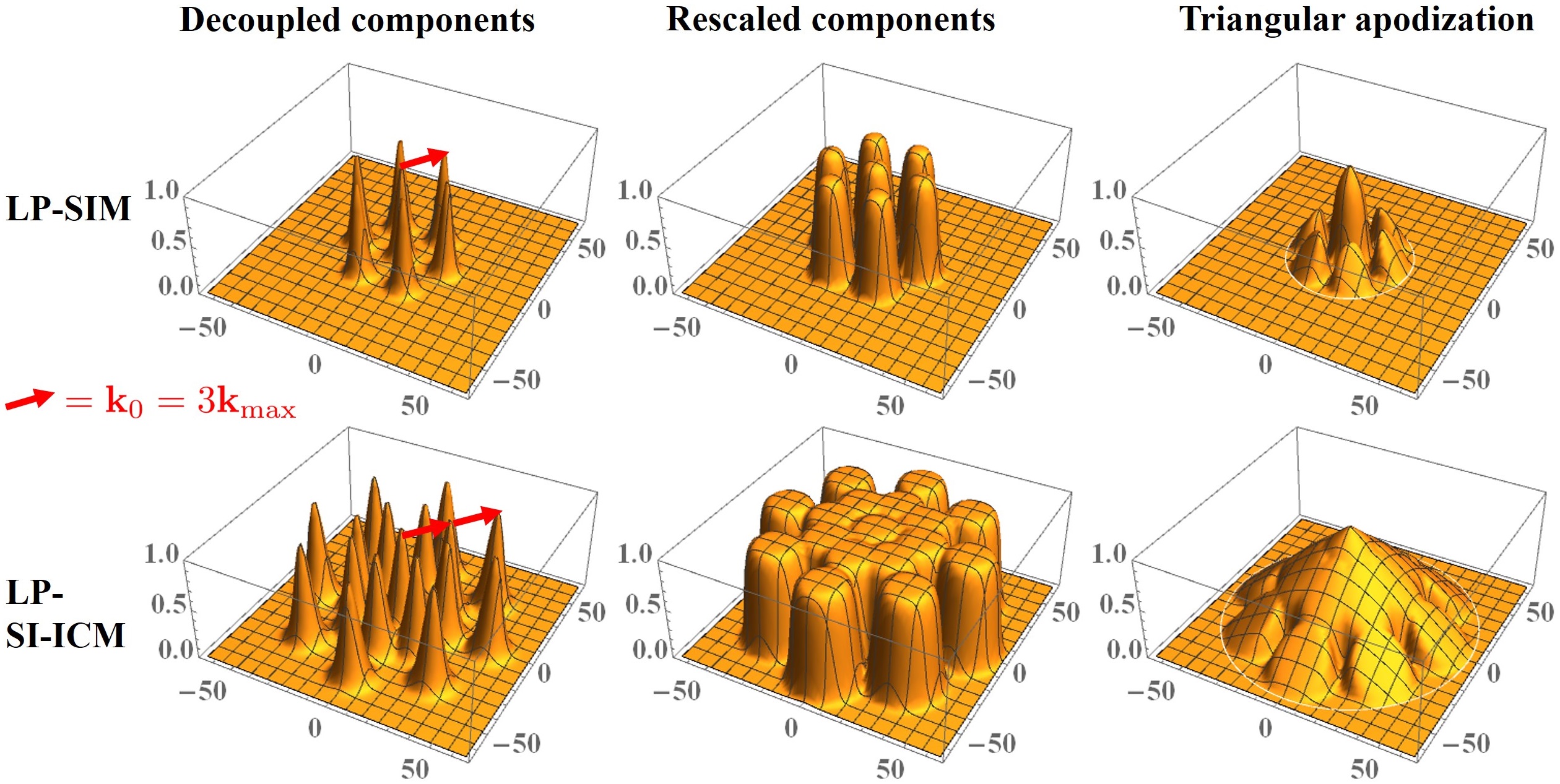}
\caption{The surface plots represent the total OTF passband for various post-processing steps. Gaussian OTF disks have been used as an approximation to the standard 2D widefield OTF. The illumination frequency is chosen as $|\mathbf{k}_0| = 3 k_\text{max}$, creating gaps in the OTF support of SIM but not SI-ICM.}
\label{fig:1}
\end{figure}

\begin{figure*}[t]
  \centering
  \includegraphics[width=0.88 \textwidth]{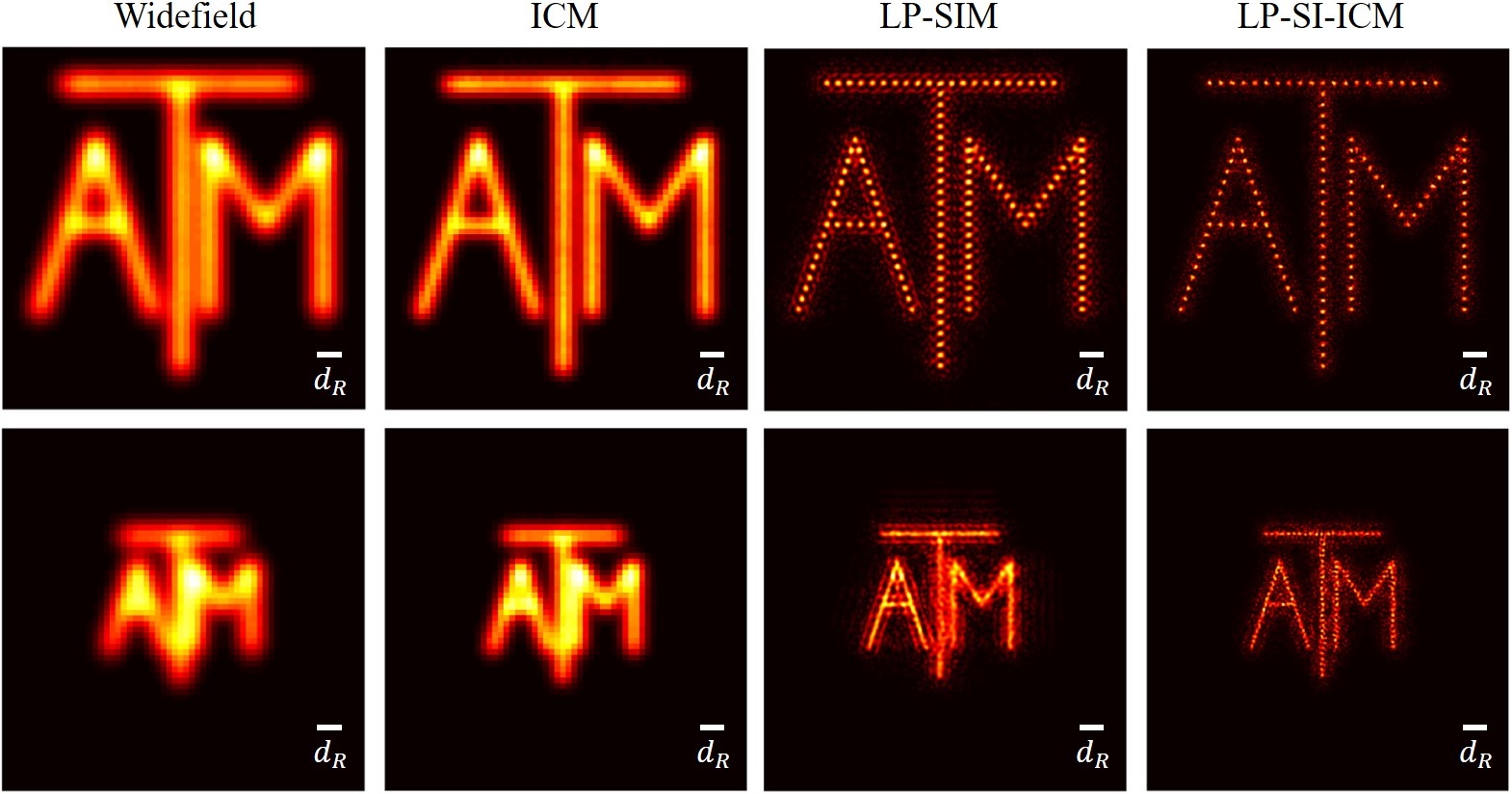}
\caption{The two samples consist of emitters aligned along the letters A, T and M. In the upper sample the emitter spacing is less than half the Rayleigh limit $d_R$, indicated by the white bar. The original pixel size corresponds to $d_R \approx 4.36$ pixels, fulfilling Nyquist sampling for widefield and ICM. The LP-SIM and LP-SI-ICM image Fourier spectra have been eight-fold zero-padded to accommodate the increased resolution in real space. The lower sample is shrunk by a factor of 2, rendering emitter separation smaller than $d_R/4$. The images show final results for widefield microscopy, ICM, LP-SIM, and LP-SI-ICM (left to right).} 
\label{fig:2}
\end{figure*}

Here, we propose to combine SI-ICM with localized plasmon SIM (LP-SIM) \cite{Ponsetto2017,Bezryadina2019}.  LP-SIM enables sub-wavelength illuminations with $\mathbf{k}_{0} > \mathbf{k}_\text{max}$,  yet $\mathbf{k}_{0} = 2\mathbf{k}_\text{max}$ should not be exceeded to prevent gaps in the OTF support (see upper row in Fig.~\ref{fig:1}). The option to work with $\mathbf{k}_{0} \geq 2\mathbf{k}_\text{max}$ and gaps in the OTF support has been explored \cite{Bezryadina2019}, but such an approach is not deterministic and can lead to artifacts since spatial frequency information in the gaps is truly lost. The gaps in the OTF support will result in a real-space PSF with strong side-lobes. While this may be OK for sparse samples, the side-lobes can significantly skew the final image in non-sparse samples (e.g. compare the upper and lower LP-SIM images in the third column of Fig.~\ref{fig:2}). Notably, the goal of superresolution imaging techniques is to resolve very fine close-by structures which inherently must carry a higher fluorophore density to be imaged with a high fidelity (according to the Nyquist sampling theorem). The deterministic resolution enhancement of LP-SIM alone is thus limited to the factor 3. LP-SI-ICM, however, would highly benefit from spatial frequencies $\mathbf{k}_0 > 2 \mathbf{k}_\text{max}$ since the enlarged OTF of ICM $H_2(\mathbf{k}) \equiv FT  \{ h(\mathbf{r})^2 \}$ prevents an early formation of gaps and spatial frequency mixing with the higher harmonic $\cos (2 \mathbf{k}_0 \mathbf{r})$ samples very high spatial frequency information around $\pm 2 \mathbf{k}_0$. The total OTF passband for LP-SIM and LP-SI-ICM with illumination wave vector $\mathbf{k}_0 = 3 \mathbf{k}_\text{max}$ is illustrated in Fig.~\ref{fig:1}. A mostly homogenous gap-less OTF support and a highly enhanced total passband are clearly visible.

The object to be imaged $n(\mathbf{r}) \propto \sum_{i=1}^N \delta(\mathbf{r}-\mathbf{r}_i)$ is assumed to be an ensemble of point-like emitters at positions $\mathbf{r}_i$. We assume the illumination $I_{\text{str}}(\mathbf{r},\alpha,\varphi_j)$ to be flat for widefield and ICM, and sinusoidally modulated for SIM and SI-ICM, with $\alpha$ and $\varphi_j$ being the orientation and phase of the pattern. The mathematical expressions for the widefield microscopy, ICM, SIM, and SI-ICM signals are then given as \cite{Classen2018}. 
\begin{equation}
\begin{aligned}
 \braket{I(\mathbf{r},t)} & = \sum\nolimits_{i=1}^N h(\mathbf{r}-\mathbf{r}_i)  \\ 
\text{ICM}_2(\mathbf{r}) &  = \sum\nolimits_{i=1}^N h(\mathbf{r}-\mathbf{r}_i)^2   \\
\text{SIM}(\mathbf{r})& = \sum\nolimits_{i=1}^N  h(\mathbf{r}-\mathbf{r}_i)  \, I_{\text{str}}(\mathbf{r}_i,\alpha,\varphi_j) \\
\text{SI-ICM}_2(\mathbf{r})&  =  \sum\nolimits_{i=1}^N  h(\mathbf{r}-\mathbf{r}_i)^2  \, I_{\text{str}}(\mathbf{r}_i,\alpha,\varphi_j)^2 \,  
\end{aligned}
\label{eq:formulas}
\end{equation}
Note that, the second and fourth line are the result of an intensity correlations analysis conducted on a set of raw frames that capture the blinking statistics of the fluorophores \cite{Dertinger2009}, or the result of antibunching photon-correlation measurements \cite{Schwartz2013,Tenne2019}. Taking the Fourier transforms yields.
\begin{equation}
\begin{aligned}
  FT \{ \text{SIM}(\mathbf{r}) \} & = H(\mathbf{k}) \times \sum\nolimits_{j} c_j \, e^{i  \varphi_j }  \,  \tilde{n}(\mathbf{k} - \mathbf{k}_j) \\  FT  \{ \text{SI-ICM}_2(\mathbf{r}) \} & =  H_2(\mathbf{k}) \times \sum\nolimits_{l}  d_l \, e^{i  \phi_l }  \,  \tilde{n}(\mathbf{k} - \mathbf{k}_l) 
\end{aligned}
\label{eq:1}
\end{equation}
where the sums are taken over the spatial frequency components $(\mathbf{k}_j, \varphi_j, c_j)$ and $(\mathbf{k}_l, \phi_l, d_l)$ contained in $I_\text{str}(\mathbf{r})$ and $I_\text{str}(\mathbf{r})^2$, respectively. Note that, the enlarged OTF $H_2(\mathbf{k})$ also reduces the need for many orientations $\alpha$ (as compared to non-linear SIM techniques \cite{Gustafsson2005}). 

The individual components can be computationally unmixed from a set of linearly independent images, which are acquired by varying the phases $\varphi_j$ and $\phi_l$, respectively. Post-processing is achieved by shifting the disentangled components back to their true positions in Fourier space [$c_j H(\mathbf{k}) \tilde{n}(\mathbf{k} - \mathbf{k}_j) \rightarrow  \tilde{n}_j(\mathbf{k}) \equiv c_j  H(\mathbf{k} + \mathbf{k}_j) \tilde{n}(\mathbf{k}) $], re-scaling and appropriate merging into a highly enlarged OTF support (see Fig.~\ref{fig:1}), mathematically expressed through the formula \cite{Gustafsson2008}
\begin{equation}
\tilde{n}_\text{new}(\mathbf{k}) = \frac{\sum_j c_j^\ast H^\ast(\mathbf{k}+\mathbf{k}_j)  \tilde{n}_\text{j}(\mathbf{k})}{\sum_j | c_j \, H(\mathbf{k}+\mathbf{k}_j) |^2 + \gamma^2} A(\mathbf{k}) \, ,
\label{eq:SIM1}
\end{equation}
where $\gamma^2$ is a constant that prevents division by zero and should be chosen noise-dependently. $A(\mathbf{k})$ is a triangular apodization function which reduces ringing in the final real-space image. The maximum cut-off values for SIM and SI-ICM are $k_\text{max} + k_0$ and $2 (k_\text{max} + k_0)$, respectively. If noise in the outer rims of the OTF support is too strong the cut-off values can be slightly reduced to suppress noise. 
Finally, we point out that any plasmon assisted illumination leads to intrinsic surface imaging since the electric field distribution $E = \exp(i k_x x - \sigma z)$ with $\sigma^2 = - (2 \pi / \lambda )^2 + k_x^2$ decays along the $z$-axis, thus imaging structures that are located within 100~nm of the substrate surface \cite{Ponsetto2017}.

Simulations of final images for widefield microscopy, ICM, LP-SIM, and LP-SI-ICM are shown in Fig.~\ref{fig:2}, and strikingly illustrate the potential advantage of combining LP-SIM with the SI-ICM principle. Here we assumed sequential sinusoidal illumination patterns in a 2D plane. We utilized three different software tools. The simulation of raw frames with blinking fluorophores was implemented via the SOFI Simulation Tool (SOFIsim) software described in \cite{Girsault2016}.Note that Poisson sampling of realistic photon number counts into the CCD pixel array intrinsically accounts for shot noise in the raw frames. External and readout noise can be added along with a finite quantum efficiency of the CCD camera. We modified the software to include the user-defined ``ATM'' mask and to enable scaling of emitter intensities by $I_{\text{str}}(\mathbf{r}_i,\alpha,\varphi_j)$, as in the third line of Eq.~(\ref{eq:formulas}). The auto-correlation analysis to obtain intermediate ICM frames was conducted via the Fourier SOFI (fSOFI) software package described in \cite{Stein2015}. The intermediate LP-SIM frames are the averages of the simulated raw frames per orientation and phase (without correlation analysis). The final widefield and ICM images are simply the sum over all orientations and phases of the intermediate LP-SIM and ICM ``frames'', respectively. The intermediate LP-SIM and ICM frames were post-processed according to eqs.~(\ref{eq:1}) and (\ref{eq:SIM1}) by custom-written software in Matlab. If the patterns are subject to aberrations or are in general unknown the reconstruction can still be achieved by use of so-called Blind SIM algorithms. These algorithms are robust and were previously used in LP-SIM since LP patterns are not perfectly sinusoidal \cite{Ponsetto2017,Bezryadina2019}. The details of these procedures can be looked up elsewhere \cite{Gustafsson2000,Gustafsson2008,Classen2017a,Classen2018,Bezryadina2019,Ayuk2013,Labouesse2016}. Other experimental imperfections (such as aberrations in the detection PSF) may reduce the SNR but would not render the technique unviable in general.

The simulation parameters for the optical system were chosen as: wavelength $\lambda = 600\,$nm, magnification $M = 100$, numerical aperture $\mathcal{A} = 1.3$, pixel size $6.45\,\mu m \times 6.45\,\mu m$, and an array of $ 64 \times 64$ pixels. The simulation software utilizes a Gaussian PSF $h(x,y) = \exp[-(x^2+y^2)/2\sigma^2]$ to approximate the well-known Airy disk $h(r) = (2 J_1(r)/r)^2$. The FWHM of the Gaussian PSF is chosen according to the Rayleigh resolution limit $d_R \equiv 0.61 \lambda/\mathcal{A}$ to roughly match the resolution power of the Airy disk. Note however, that it might be more realistic to match the respective FWHMs of both PSFs, which for the Airy disk reads $\approx 0.51 \lambda/\mathcal{A}$, resulting in a close to $20\,\%$ discrepancy. In our SIM software we kept the original $d_R \equiv 0.61 \lambda/\mathcal{A}$ scaling for the Gaussian PSF and OTF, respectively, to not alter the reconstruction. For the illumination we chose $k_0 = 2 \, k_\text{max}$. Accounting for the discrepancy in the FWHM of the PSFs one can argue that is effectively closer to the value $k_0 \approx 2.4 \, k_\text{max}$. Before final assembly, the LP-SIM and LP-SI-ICM image Fourier spectra have been eight-fold zero-padded to accommodate the increased resolution in real space. The original pixel size corresponds to $d_R \approx 4.36$ pixels, fulfilling Nyquist sampling for the widefield and ICM images.

In the simulation software the frame rate was set as 100 frames/s (10~ms per frame). The acquisition time per orientation and phase of the structured illumination pattern was set to 2~s, resulting in 200 raw frames each. The fluorophore on- and off-time were set to 20~ms and 40~ms, respectively. The photon number per emitter (in the on state) per frame was set to 800. The background photon count was set to zero. In addition the camera gain was set to 6 and the quantum efficiency to 0.9, the readout noise to the lowest value allowed by the software 0.1~rms and the dark current to the lowest value $0.001$ electrons/pixel/s. We call this setting the ``low noise/strong signal'' scenario. 

The larger ATM structure is just resolved by widefield microscopy and slightly better resolved by the ICM signal (improvement of $\sqrt{2}$, without deconvolution). The individual emitters are not resolved since their separation is smaller than half the Rayleigh limit $d_R /2$. LP-SIM and LP-SI-ICM resolve each emitter position, but the LP-SIM PSF suffers from strong side-lobes due to small gaps in the OTF support (due to $k_0 \approx 2.4 \, k_{max}$). 
The second structure is two-fold shrunk in size and hardly resolved by the widefield and ICM images. LP-SIM achieves a higher resolution but the strong side-lobes of the LP-SIM PSF reduce the image fidelity, making the reconstruction not fully deterministic. LP-SI-ICM is still capable of resolving the structure with a high fidelity and some individual emitter positions, but in general the influence of (shot) noise is already apparent. The advantages of LP-SI-ICM are two-fold. First, the second harmonic reaches out to  $\pm 2 |\mathbf{k}_0| \approx \pm 4.8\, k_\text{max}$ and second, the individual OTF disks $H_2(\mathbf{k}) \equiv FT  \{ h(\mathbf{r})^2 \}$ prevent an early formation of gaps. Note that for LP-SIM utilizing the theory PSF/OTF without noise would not significantly enhance the image fidelity since the strong PSF side-lobes can not be removed. The value $k_0 \approx 2.4 \, k_{max}$ for LP-SI-ICM still highly suppresses the appearance of side-lobes and this value can be readily implemented with previous experimental setups and plasmonic substrates made of silver nano-pillars of 60 nm size and $100-200$ nm pitch \cite{Ponsetto2017,Bezryadina2019}. Varying the wavelength and/or the numerical aperture of the microscope objective for fluorescence collection would change the effective value of $k_0$.

\begin{figure}[t]
  \centering
  \includegraphics[width=0.88 \columnwidth]{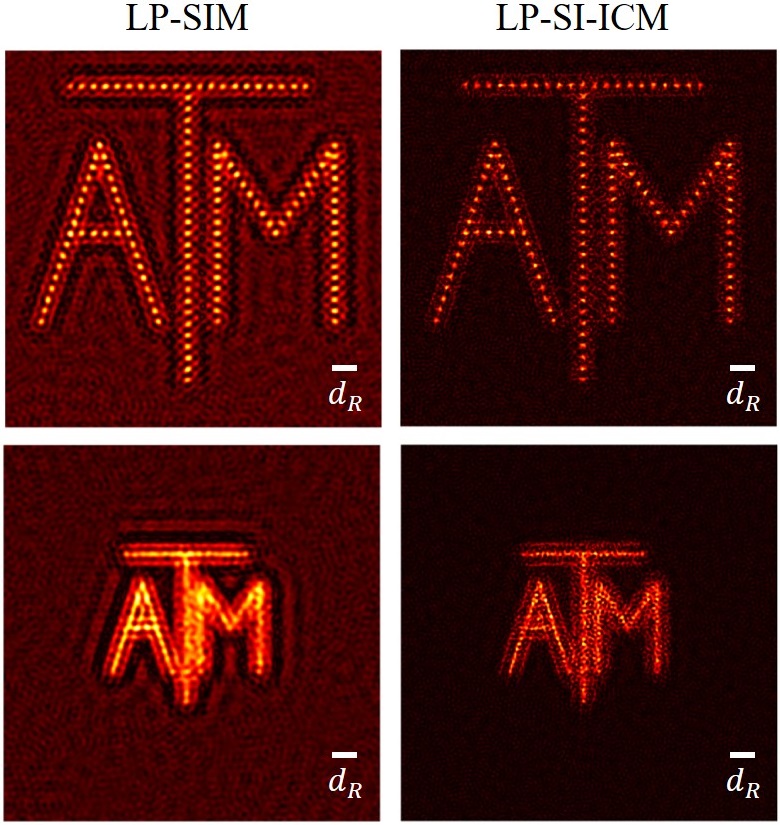}
\caption{The samples are equivalent to Fig.~\ref{fig:2}. In the simulation the ``high noise/low signal'' settings were used (see text).}
\label{fig:3}
\end{figure}

The image acquisition procedure would require a total of 50 s (10 ms/frame, 25 patterns, 200 raw frames). Adjusting the phase and orientation of the structured illumination patterns can be implemented with fast mirror galvanometers such that we do not assign any time. Post processing of the data can be executed later. If necessary the software can be adjusted to provide a close-to live feedback, since the intensity correlation analysis and the SIM post-processing are fast. To achieve faster imaging a higher frame rate on the order of 1 ms/frame would be beneficial. The number of orientations for the structured illumination can be reduced to three, and the raw frames per pattern to 100. The total measurement time would then read 1.5 s (1 ms/frame, 15 patterns, 100 raw frames). To counterbalance lower photon counts the illumination intensity should be increased.

To illustrate a reduction in raw frames and lower signal levels, an additional simulation was conducted for ``high noise/low signal'' settings, which were chosen as 100 raw frames per illumination pattern, 200 photons per emitter (in the on state) per frame, 2 background photons per camera pixel per frame, 0.7 quantum efficiency, 6 camera gain, 1.6 rms pixel readout noise, 0.06 electrons/pixel/s dark current. In the reconstruction we slightly reduced the size of the triangular cut-off and increased the size of $\gamma$ to suppress noise (see Eq.~(\ref{eq:SIM1})).  The resulting final images for the same ``ATM'' structures are shown in Fig.~\ref{fig:3}. The final images suffer from enhanced noise and thus represent lower fidelity reconstructions. Nonetheless the basic structure is still apparent, both for LP-SIM and LP-SI-ICM. 


To quantify the resolution further we conducted some additional simulations with randomly distributed emitters and applied a FRC analysis \cite{Banterle2013}. We used standard settings and looked at the 1/2-bit and 1-bit resolution criteria. We obtained the following resolution values from low noise simulations of WF, ICM, LP-SIM and LP-SI-ICM images (from left to right, in units of pixels): (25.0, 18.2, 9.0, 4.5)  for 1/2-bit, and (25.0, 19.2, 9.1, 5.6) for 1-bit. Note that the FRC analysis yields slightly better results for WF microscopy than expected from the Rayleigh limit $8 \times 4.36\,\text{pixels} \approx 34.5\,\text{pixels}$ (after interpolation). For high noise simulations we obtained (from left to right, in units of pixels): (26.3, 22.2, 9.7, 5.9)  for 1/2-bit, and (27.0, 23.2, 9.8, 6.5) for 1-bit. The results are generally in good agreement with the visual perception of the resolution enhancements. 

Another avenue to significantly imaging speeds would be the use of antibunching correlation measurements \cite{Classen2017a,Tenne2019,Rossman2019}. This might be possible in the future with novel camera technology such as SPAD arrays. The SNR values for ultra short measurement times can then be quite low, but one can utilize the same imaging speed as for LP-SIM. Moreover, the lower resolution SIM data (from the same data set) can be augmented by the high-resolution low-SNR quantum correlation data via joint sparse recovery, as has been demonstrated in the confocal variant of SI-ICM with antibunching correlation measurements \cite{Rossman2019}. 

In conclusion, we proposed to enhance localized plasmon SIM by use of intensity auto-correlations. The advantages are two-fold. First, the resolution is highly enhanced because higher harmonics of the illumination pattern arise and second, the larger OTF from intensity correlations prevents an early formation of gaps in the OTF support. We presented simulations that demonstrate the performance for realistic photon budgets and with external noise sources. With the development of camera technologies to reliably measure antibunching photon correlations the imaging speed may be highly enhanced. The proposed technique has potential in light microscopy where low-intensity illumination is paramount while aiming for high spatial but moderate temporal resolutions. 

\section*{Funding Information}

Robert A. Welch Foundation (Award numbers: A-1943-20180324 and A-1801-20180324);
Russian Science Foundation (project 20-12-00088 – ultrabroadband optics);
Ministry of Science and Higher Education of the Russian Federation (14.Z50.31.0040-17/02/2017);
Russian Foundation for Basic Research (projects 18-29-20031, 19-02-00473);

\section*{Acknowledgements}

A. C. acknowledges support from the Alexander von Humboldt Foundation in the framework of a Feodor Lynen Research Fellowship.

\section*{Disclosures}

The authors declare no conflicts of interest.


\begin{thebibliography}{10}
\newcommand{\enquote}[1]{``#1''}

\bibitem{Gustafsson2000}
M.~G.~L. Gustafsson, {\protect\JournalTitle{J. Microsc.}} \textbf{198}, 82
  (2000).

\bibitem{Hell1994}
S.~W. Hell and J.~Wichmann, {\protect\JournalTitle{Opt. Lett.}} \textbf{19},
  780 (1994).

\bibitem{Betzig2006}
E.~Betzig, G.~H. Patterson, R.~Sougrat, O.~W. Lindwasser, S.~Olenych, J.~S.
  Bonifacino, M.~W. Davidson, J.~Lippincott-Schwartz, and H.~F. Hess,
  {\protect\JournalTitle{Science}} \textbf{313}, 1642 (2006).

\bibitem{Dertinger2009}
T.~Dertinger, R.~Colyer, G.~Iyer, S.~Weiss, and J.~Enderlein,
  {\protect\JournalTitle{Proc. Natl. Acad. Sci. USA}} \textbf{106}, 22287
  (2009).

\bibitem{Schwartz2013}
O.~Schwartz, J.~M. Levitt, R.~Tenne, S.~Itzhakov, Z.~Deutsch, and D.~Oron,
  {\protect\JournalTitle{Nano Lett.}} \textbf{13}, 5832 (2013).

\bibitem{Strohl2016}
F.~Str\"{o}hl and C.~F. Kaminski, {\protect\JournalTitle{Optica}} \textbf{3},
  667 (2016).

\bibitem{Classen2017a}
A.~Classen, J.~von Zanthier, M.~O. Scully, and G.~S. Agarwal,
  {\protect\JournalTitle{Optica}} \textbf{4}, 580 (2017).

\bibitem{Classen2018}
A.~Classen, J.~von Zanthier, and G.~S. Agarwal, {\protect\JournalTitle{Opt.
  Express}} \textbf{26}, 27492 (2018).

\bibitem{Tenne2019}
R.~Tenne, U.~Rossman, B.~Rephael, Y.~Israel, A.~Krupinski-Ptaszek,
  R.~Lapkiewicz, Y.~Silberberg, and D.~Oron, {\protect\JournalTitle{Nat.
  Photon.}} \textbf{13}, 116 (2019).

\bibitem{Sroda2020}
A.~Sroda, A.~Makowski, R.~Tenne, U.~Rossman, G.~Lubin, D.~Oron, and
  R.~Lapkiewicz, {\protect\JournalTitle{Optica}} \textbf{7}, 1308 (2020).

\bibitem{Ponsetto2017}
J.~L. Ponsetto, A.~Bezryadina, F.~Wei, K.~Onishi, H.~Shen, E.~Huang,
  L.~Ferrari, Q.~Ma, Y.~Zou, and Z.~Liu, {\protect\JournalTitle{ACS Nano}}
  \textbf{11}, 5344 (2017).

\bibitem{Bezryadina2019}
A.~Bezryadina, J.~Zhao, Y.~Xia, Y.~U. Lee, X.~Zhang, and Z.~Liu,
  {\protect\JournalTitle{Opt. Lett.}} \textbf{44}, 2915 (2019).

\bibitem{Gustafsson2005}
M.~G.~L. Gustafsson, {\protect\JournalTitle{Proc. Natl. Acad. Sci. USA}}
  \textbf{102}, 13081 (2005).

\bibitem{Gustafsson2008}
M.~G.~L. Gustafsson, L.~Shao, P.~M. Carlton, C.~J.~R. Wang, I.~N. Golubovskaya,
  W.~Z. Cande, D.~A. Agard, and J.~W. Sedat, {\protect\JournalTitle{Biophys.
  J.}} \textbf{94}, 4957  (2008).

\bibitem{Girsault2016}
A.~Girsault, T.~Lukes, A.~Sharipov, S.~Geissbuehler, M.~Leutenegger,
  W.~Vandenberg, P.~Dedecker, J.~Hofkens, and T.~Lasser,
  {\protect\JournalTitle{PLOS ONE}} \textbf{11}, 1 (2016).

\bibitem{Stein2015}
S.~C. Stein, A.~Huss, D.~H\"{a}hnel, I.~Gregor, and J.~Enderlein,
  {\protect\JournalTitle{Opt. Express}} \textbf{23}, 16154 (2015).

\bibitem{Ayuk2013}
R.~Ayuk, H.~Giovannini, A.~Jost, E.~Mudry, J.~Girard, T.~Mangeat, N.~Sandeau,
  R.~Heintzmann, K.~Wicker, K.~Belkebir, and A.~Sentenac,
  {\protect\JournalTitle{Opt. Lett.}} \textbf{38}, 4723 (2013).

\bibitem{Labouesse2016}
S.~{Labouesse}, M.~{Allain}, J.~{Idier}, S.~{Bourguignon}, A.~{Negash},
  P.~{Liu}, and A.~{Sentenac}, \enquote{Fluorescence blind structured
  illumination microscopy: A new reconstruction strategy,} in \emph{2016 IEEE
  International Conference on Image Processing (ICIP),}  (2016), pp.
  3166--3170.

\bibitem{Banterle2013}
N.~Banterle, K.~H. Bui, E.~A. Lemke, and M.~Beck, {\protect\JournalTitle{J.
  Struct. Biol}} \textbf{183}, 363 (2013).

\bibitem{Rossman2019}
U.~Rossman, R.~Tenne, O.~Solomon, I.~Kaplan-Ashiri, T.~Dadosh, Y.~C. Eldar, and
  D.~Oron, {\protect\JournalTitle{Optica}} \textbf{6}, 1290 (2019).

\end{thebibliography}

\end{document}